\documentclass[12pt]{article}

\begin{document}

\begin{flushright}
CERN-TH/2000-341 \\
IITAP-2000-007 \\
hep-th/0011272 \\
November  2000
\end{flushright}
\vspace*{0.7cm}
\begin{center}
{\bf THE CHALLENGE OF LIGHT-FRONT QUANTISATION: \\
RECENT RESULTS}
\footnote{Based on the lecture delivered by GBP at 
the XXXIV PNPI Winter School, Repino, St.Petersburg, Russia, 
February 14-20, 2000, to appear in the Proceedings.}\\ 
\vspace*{1cm}
{ \bf Victor~T.~Kim${}^{\ddagger \&}$, Victor~A.~Matveev${}^{\S}$,
Grigorii~B.~Pivovarov${}^{\S }$ \\
{\rm and} \\
James~P.~Vary${}^{\dagger}$}  \\
\vspace*{0.7cm}
\end{center}
${}^\ddagger$ : St.Petersburg Nuclear Physics Institute,
188300 Gatchina, Russia
\newline
${}^\&$ : CERN, CH-1211, Geneva 23, Switzerland
\newline
${}^\S$ : Institute for Nuclear Research, 117312 Moscow, Russia
\newline
${}^\dagger$ : Department of Physics and Astronomy,
Iowa State University, \\ 
\hspace*{0.45cm} Ames, IA 50011, USA

\begin{center}
Abstract
\end{center}

We explain what is the challenge of light-front quantisation, 
and how we can now answer it because of recent progress 
in solving the problem of zero modes in the case of non-Abelian 
gauge theories. We also give a description of the 
light-front Hamiltonian for $SU(2)$ finite volume gluodynamics 
resulting from this recent solution to the problem of 
light-front zero modes.

\vspace{1cm}
\newpage

{\bf 1. Introduction: What is the challenge of 
light-front quantisation? }\\

The idea of light-front quantisation (or {\it light-cone} quantisation) 
of  relativistic dynamical systems is more than fifty years old. 
It was introduced by Dirac \cite{Dirac}. In a nutshell, it suggests 
taking a point of view of a massless observer 
flying with the speed of light. 
The picture of the relativistic dynamics such an observer would have 
is much different from the 
conventional one. In the most general terms, the challenge of light-front 
quantisation is to exploit the possible advantages of this  
light-like observer's point of view.

Light-front quantisation has been extensively studied and \
used for fifty years. For a recent review, 
see Ref. \cite{Brodsky}.
Despite those massive efforts, the challenge is still not 
answered properly. To explain why, we first explain what is 
potentially the most prominent advantage 
of the light-front description.

The major complication of a complete (nonperturbative) treatment in the 
conventional equal-time quantisation of 
relativistic dynamical systems (e.g., Lorentz invariant field theories) 
is that it is not possible to single out a finite "set of characters" taking 
part in the events. If we single out some set of particles, 
new particles are created through interactions, and 
the number of particles is always potentially infinite.
At the moment, the only practical way out of that intractable 
situation is the lattice regularisation. 
Under that regularisation, the situation is put under control by 
keeping the number of degrees of freedom in the description 
proportional to the 
number of the vertices of the lattice that replaces
the continuum space. The biggest promise of the light-front 
quantisation is that it 
may give an alternative way to overcome the problem of 
an infinite number of 
interacting degrees of freedom. Therefore, the challenge of 
light-front quantisation is to develop it to a stage where 
it would be a serious rival to the lattice field theory. 
Evidently, it is nowhere close to that stage of
development.

The way the light-front keeps the infinite 
number of interacting 
degrees of freedom under control is very different from lattice 
regularisation. It is less severe: the initial theory 
is distorted
not at small and large distances, as is the case with 
the lattice regularisation, 
but only at large distances. Because of that, the total 
number of degrees of freedom
the light-front formulation deals with is infinite, in sharp contrast to the
lattice regularisation. Instead of cutting the total number of degrees of 
freedom, light-front quantisation gives a possibility to break the total 
set of degrees of freedom into 
subsets, each of the 
subsets finite, in such a way that
there would be no interaction between the degrees of freedom from different 
subsets. One may say that the light-front promises to slice 
the complete theory into an
infinite set of independent subtheories in such a way that 
each subtheory constitutes 
a quantum mechanics with a finite number of degrees of freedom.

Each subtheory is singled out by a value of an additive 
semi-positive conserving 
quantum number. The availability of such a quantum number is 
the major characteristic feature of the light-front quantisation. 
Let us explain what 
this quantum number is.

Let the light-like observer 
be flying along the third axis 
in the positive direction. 
The natural "time" (i.e., the coordinate 
parametrising the world-line 
of the observer) is then $x^+\equiv (x^0+x^3)/\sqrt{2}$
(the normalising square root above is a matter of later convenience).
Therefore, the hyperplane in the space-time where the light-like 
observer is setting the 
initial conditions for all the dynamics is singled out by 
the condition $x^+=const$. 
In analogy with the equal-time formulation,
the components of the momentum generating the shifts of the system at
fixed $x^+$ are kinematical, i.e., even in the presence of 
an interaction between the parts of the system, the total momentum 
is a sum of the momenta of the subsystems. 
In other words, the operators corresponding to the kinematical 
variables are quadratic in the creation--annihilation operators 
and easy to diagonalise  even in the 
presence of an interaction. To reiterate, the components 
of the momentum $P_- = (P_0 - P_3)/\sqrt{2}$, $P_\perp$ are 
kinematical in the light-front quantisation
($P_\perp$ denotes the set of the space-like momentum components 
perpendicular to the third direction). 
This is quite similar to the kinematical character of 
$P_3$ and $P_\perp$ in the equal-time quantisation. 
What is not similar is the fact that 
$P_-$ is {\it non-negative} 
($P_- = (P_\perp^2 + M^2)/(2P_+)$, 
where $M$ is the mass of the state). 

This qualitative difference between the equal-time and 
light-front quantisations 
enables the following consideration: Suppose we make an infrared 
regularisation of the system that discards all excitations whose 
$P_-$ is smaller than some 
regularisation parameter, and let $P_-$ 
be conserving for 
the regularised system as 
it is for the original system. Then the consideration of the dynamics 
can be restricted to the sectors of fixed 
$P_-$, and  
every such sector cannot accommodate 
more excitations than its $P_-$ divided by the minimal $P_-$ allowed. 
That is the case because adding one more excitation would 
increase $P_-$ at least by the 
minimal value allowed for $P_-$. Notice 
that it is not the case for 
the equal-time quantisation: because the negative values of momentum 
are allowed in that case, it is always possible to add more 
excitations to any state without changing its momentum.

We summarise  that the promise of the light-front quantisation 
is high: it promises 
to give a non-perturbative 
definition of quantum field theories 
that may compete and complement the lattice formulation.

In what follows, we will see how this promise fails in general, and how the 
non-Abelian gauge theory escapes this failure 
because of recent findings.
The rest of the paper is organised as follows. 
Section 2 discusses the problem of zero modes, 
which is the biggest threat to the light-front division into finite sectors; 
Section 3 specifies the discussion to non-Abelian gauge theories; 
Section 4 describes the 
recent findings restoring the hope for
realising of the light-front separation into sectors; 
Section 5 contains a description of the Hamiltonian 
of $SU(2)$ gluodynamics reduced to the smallest 
non-trivial sector of fixed $P_-$; 
Section 5 contains the conclusion and outlook.\\

{\bf  2. The Zero Modes }\\

It is time to specify the infrared regularisation needed 
to cut $P_-$ from below. It 
should respect the symmetries of the theory, in particular, 
the gauge symmetry
(as for the lattice formulation, the major application for 
the light-front quantisation 
should be to the non-Abelian gauge theories). The only evident 
way to achieve this 
is to compactify the $x^-$ direction. Therefore, in what follows we consider 
all the fields to be periodic in the $x^-$ direction: 
$A(x^- = L/2) = A(x^- = - L/2)$.
In that case, the spectrum of $P_-$ is discrete, and the smallest 
possible non-zero value of $P_-$ is $2\pi/L$. This approach, 
with finite $x^-$ span and discrete
$P_-$ is known as {\it discretised light-cone quantisation} (DLCQ) 
\cite{Pauli,Maskawa}.

What about the degrees of freedom that have zero $P_-$? Potentially they 
endanger our program, because it may be possible to add any number of such 
excitations to any state without changing its $P_-$. Such degrees of freedom 
correspond to the field configurations independent of $x^-$, $\partial_-A = 0$. 
This is the infamous problem of light-front zero modes. Because of it, the promise 
of light-front quantisation has yet to be realised.

The problem of the light-front zero modes was first analysed in 
Ref. \cite{Maskawa}.
It was pointed out that, by Lorentz invariance, 
the time derivative of a field $A$ 
may enter the action in the combination $\partial_+ A\partial_- A$. 
Therefore, in the cases where the time derivative enters in this combination alone,
no time derivative of the zero modes enter the action, 
because their $\partial_-$ is zero.
Thus, varying the action over the zero modes gives an equation without time 
derivatives. In other words, the zero modes are not among the dynamical 
degrees of freedom.
The classical equations of motion allow 
one to express
them in terms of the real dynamical
degrees of freedom at any moment of time, without referring 
to dynamical evolution. 
After that, substituting them back into the action yields 
a theory without zero 
modes, and the promise of the light-front quantisation survives. 
That is, it survives in principle. But practically, the equation 
for the zero modes is a nonlinear equation in partial derivatives. 
We do not know how to solve it. It is instructive to 
analyse an example of the $\phi^4$ theory to see what is the characteristic 
appearance of the zero modes equation (to obtain it, integrate the equation 
for the field over $x^-$).

We conclude that in general the light-front quantisation is stuck at 
the problem of zero modes. It remains to consider particular cases: 
what if in a case of interest the 
zero mode problem can be solved?

Our message is that it is indeed the case: we are lucky. 
The most interesting case 
of non-Abelian gauge theories has a zero mode equation 
which can be explicitly solved, and the Hamiltonians can be 
explicitly obtained for particular sectors of 
fixed value of $P_-$.

In what follows, we consider the simplest relevant case of $SU(2)$ 
gluodynamics.\\

{\bf 3. Light-Front Gluodynamics }\\

The first complication with gluodynamics is that in general there are time 
derivatives of the field configurations independent of $x^-$ in the action. 
That is the case because there is a four-vector apart from 
$\partial_\mu$ at our disposal 
(it is the gauge field $A_\mu$).
Evidently, combinations like $(\partial_+\phi)A_-\phi$, where 
$\phi$ is a matter 
field, or transverse components of the gauge field, are present in the action.
Their presence complicates
the zero mode issues as well as the Hamiltonian treatment, 
because for the Hamiltonian treatment 
we want the time derivatives to enter the action in the combinations 
$p_i\dot{q}_i$, where $p$ are the momenta, and $q$ are their conjugate 
coordinates.

Because of that, probably starting from Ref. \cite{Tomboulis}, 
it is customary to consider the 
light-front formulation in the light-cone gauge, $A_- = 0$. 
It is important that in 
this gauge the Euler-Lagrange equation that follows from 
variation of the action 
of gluodynamics with respect to $A_-$ is implied by the rest of the equations. 
Thus, it is consistent to set $A_- = 0$ right in the action. 
Variation over $A_+$ 
gives an equation for $A_+$ without time derivatives. The action at $A_- = 0$, 
and at $A_+$ excluded by the equations of motion is amenable to the 
Hamiltonian treatment.

It was noticed in Ref. \cite{Novozhilov} that at finite span of 
$x^-$ the light-cone gauge is inaccessible. 
This is the case because at finite volume there is a gauge invariant quantity 
depending only on $A_-$. It is the trace of the large Wilson loop 
embracing the hole 
span of the $x^-$ direction: 
$W = {\rm Tr}P\exp\big( ig\int_{-L/2}^{L/2}\,dx^-A_-\big) /2$. 
If $W$ deviates from unity, no gauge transformation can change this fact. 
The closest one can get to the light-cone gauge at finite volume 
is to keep $A_-$ 
diagonal and independent of $x^-$. If we take this gauge (we will call it 
light-cone gauge even if its $A_-$ is non-zero), 
the problem of the Hamiltonian treatment is 
non-trivial.  In particular, it was studied in Ref. \cite{Novozhilov}, 
and in a number of subsequent papers. The results of those studies 
were summarised in Ref. \cite{Prokhvat}. We extract from Ref. 
\cite{Prokhvat} the conclusion relevant for our 
consideration: the problem of zero modes is 
very complicated in the case under consideration. Other studies 
\cite{Pauli1} seem to agree with that conclusion.

Our recent work \cite{KMPV} sets the whole picture 
in a new perspective. 
The major conclusion is opposite: the zero mode equation is 
{\it linear} with respect to the zero 
modes, it is quite possible to solve it, and to write 
down explicit expressions for 
the Hamiltonians at fixed values of $P_-$.

The reason for this qualitative difference in conclusions lies 
in the choice of variables: 
in Ref. \cite{Prokhvat}, 
the formulation is made in the traditional light-cone gauge, 
while in Ref. \cite{KMPV}, 
there is no particular gauge choice, and the determination 
of the canonical variables, zero 
modes, Gauss law, etc. is made prior to any gauge choice. 
This enables a direct 
approach to solving the problem with fully retained zero modes.

In the next Section, we sketch the results of Ref. \cite{KMPV}, and then, 
in Section 5, 
give a description of the Hamiltonian in the sector of $P_- =2\pi/L$.\\

{\bf 4. Canonical Variables and Zero Modes of $SU(2)$ Gluodynamics }\\

In Ref. \cite{KMPV}, the problem of Hamiltonian treatment of the 
light-front gluodynamics has been reconsidered using 
the  Faddeev-Jackiw 
approach \cite{Jackiw} to constrained systems. In this Section, we 
give a simplified version of the treatment of Ref. \cite{KMPV}.

There are two key steps to determine the canonical 
structure of the light-
front gluodynamics \cite{KMPV}. The first step is made in analogy 
with the equal-time treatment. It is related with the determination 
of the variable canonically 
 conjugate to $A_-$. The analogy hinges on the fact that 
the field equations for 
$A_-$ have second time derivatives, which are related to 
the terms of the action 
whose structure is $\partial_+A_-\partial_+A_-$. 
Notice that in this respect $A_-$ 
is similar to all the space components of the gauge field 
if the dynamics of the 
latter is considered in the equal-time approach. Therefore, 
by this analogy, one of 
the canonical variables is $A_-$, and its canonical conjugate is 
$E=F_{+-} + \ldots$. 
The dots denote the terms whose appearance is due to 
the non-trivial dependence of the rest of the canonical 
variables on $A_-$ (see below). 
Because of that dependence, time derivatives of $A_\perp$, 
when expressed in terms of the 
canonical variables, contain terms proportional to 
the time derivatives of $A_-$, 
thus extra contributions to $E$ appear.

The second step is related to the treatment of the terms 
of the action containing 
time derivatives of $A_\perp$. They are
\begin{equation}
\label{tdp}
\partial_+A_\perp D_-A_\perp.
\end{equation}

Because of the form of the term (\ref{tdp}), 
it is natural to expand the field 
$A_\perp$ over the eigenfunctions of $D_-$:
\begin{equation}
\label{canonical}
A_\perp = B_\perp\chi_0 + 
\sum_{p>0}\bigg[\frac{\chi_p}{\sqrt{2p}}(a_p^\dagger)_\perp 
+\frac{\chi_p^\dagger}{\sqrt{2p}}(a_p)_\perp\bigg],
\end{equation}
where $D_-\chi_p = ip\chi_p$, $\chi^\dagger_p=\chi_{-p}$, 
and $p$ is the real eigenvalue of $D_-/i$. 
To avoid misunderstanding, $\chi_p$ 
on the right-hand side
of Eq. (\ref{canonical}) are matrices like $A_\perp$ on the 
left-hand side; $B_\perp$ 
is a real field independent of $x^-$, and 
$(a_p^\dagger)_\perp$, $(a_p)_\perp$ are complex, 
conjugated to one another, independent of $x^-$. 
There is a scalar product with 
respect to which $D_-/i$ is a Hermitian operator, 
and $\chi_p$ are its 
eigenfunctions (see Refs. \cite{KMPV,Naus}).

If this expansion of $A_\perp$ is substituted in the action, 
the obtained form of the 
action shows that $a^\dagger_\perp$, $a_\perp$ 
are the creation-annihilation 
operators, while $B_\perp$ is the zero mode (there 
are no time derivatives of 
$B_\perp$ in the action). Simultaneously, extra terms 
in $E$ are generated replacing 
the dots above (see Ref. \cite{KMPV} for details).

The good news is that the action is quadratic in $B_\perp$, 
and the equation for the zero modes is a 
{\it non-singular linear} equation. 
Yet there is one more component of $A_\mu$ which we have not 
mentioned: the $A_+$ component. There are no time derivatives 
of that component in the action, 
and the action is linear in it. This is quite similar 
to the way the $A_0$ component 
enters the action in the case of the equal-time treatment. 
Also similar is its role: 
variation over it yields the Gauss law. 
It is crucial that the Gauss law does not 
contain $B_\perp$, and the equation for $B_\perp$ 
does not contain $A_+$: they 
do not meet in the action. 

So, the general structure of the light-front theory 
is quite similar to the general 
structure of the equal-time theory: there are canonical variables determined 
regardless of the gauge (like the canonical pair 
$\bf E$, $\bf A$ in the equal-time
theory), and there is a Gauss law. The only complication 
is that we need to solve for $B_\perp$. 

Now turn back to our program: Does the promise of 
the light-front formulation persist? 
Can we slice the theory into independent quantum mechanical sectors, 
each with 
finite number of the degrees of freedom? The answer 
is in the affirmative, but there are subtleties. 
To see them, we need to give a description of the 
excitations we have in the formulation, in particular, 
to trace how the total $P_-$ 
is built up from contributions of the separate excitations.

First of all, using the Gauss law, we can demonstrate that 
only a diagonal part of $A_-$, independent of the location 
on the transverse plane and of $x^-$, is a true 
dynamical quantity. This is so because the 
non-diagonal, or $x^-$-dependent part of $A_-$ 
can be removed by a gauge 
transformation, and a diagonal part 
dependent on the location on the transverse plane can 
be expressed in terms of the 
rest of the dynamical variables through the Gauss law. 
The global contribution to $A_-$ does not contribute to $P_-$. 
Therefore, restricting $P_-$ does not restrict the dynamical 
component of $A_-$. Thus, the wave functions of the 
sectors with fixed $P_-$ will depend on this variable. 
It is important that this is 
not a field but a number variable, i.e., there is only 
one degree of freedom related 
to this variable. 
It is also important that the wave functions 
should be periodic with 
respect to this variable because of the presence of "large" 
gauge transformations 
(see Refs. \cite{KMPV,Martinovic} for details). Thus, $A_-$ does 
not threaten our 
program of slicing the theory into quantum mechanics.

Next are the "transverse gluons" excited by $(a_p^\dagger)_\perp$. 
First of all, to really count them, we need to describe 
the spectrum $p$ of $D_-$. Generally, 
each $p$ is a function of location on the transverse plane, 
and a functional of 
$A_-$. It is not very convenient to count degrees of freedom by a function
(not to say about a functional), but we really do not need to 
do this because we can 
use the natural ordering of $p$: for $SU(2)$, 
we can describe the positive part of 
the spectrum by two 
parameters;  one is a 
non-negative integer $n$, and another is 
a variable $\sigma = -1, 0, +1$. 
Then $n$ will determine $2\pi n/L$ minimising 
the difference $|p-2\pi n/L|$, and $\sigma$ 
will determine the sign of the 
deviation of $p$ off $2\pi n/L$. With such a definition, 
we gain the possibility to 
number the modes by $n$ and $\sigma$. Note that when $n=0$ 
only $\sigma=+1$ is available, because 
$(a_{0,-})^\dagger = a_{0,+}$, and the term of Eq.
(\ref{canonical}) corresponding to $n = 0$, $\sigma =0$ 
is the zero mode $B_\perp$. 
It is important to note that $p_{n,\sigma}$ 
is a non-smooth functional of $A_-$:
\begin{equation}
\label{p1}
p_{n,\sigma} = \frac{2\pi n}{L} + \sigma{\rm Dev}(gA_-),
\end{equation}
where ${\rm Dev}(*)$ is a function whose value 
is equal to the smallest absolute deviation of its argument off 
an integer multiple of $2\pi/L$. Therefore, in the 
Hamiltonian, we expect a non-smooth dependence on $A_-$ (see below).

After we know the numbering of the excitations, it is time 
to ask what is the contribution of the excitations 
to the total $P_-$ of the system. Using explicit 
expressions for $P_-$ \cite{KMPV} it is easy to verify that 
the contribution of the $(n, \sigma)$ excitation to 
$P_-$ is $2\pi n/L$. Therefore, there is an excitation 
whose contribution to $P_-$ vanishes: 
it is the $(0, +)$ excitation.

That endangers our program: there is a zero-momentum
physical degree of freedom. But the general idea of 
the light-front split survives. This is the case 
because $\sigma$ coincides with an Abelian charge 
of the excitations, and there is 
a component of the Gauss law requesting the total Abelian charge on the 
transverse plane to vanish all the time. So, if we add 
to an admissible state a single excitation $(0, +)$, 
we have to add one more excitation whose $\sigma = -1$, 
to keep the total charge zero. And the smallest 
$P_-$ of such an excitation is 
$2\pi/L$ (this is the momentum of the $(1, -)$ excitation).

We conclude that the combination of the Gauss law and 
zero mode analysis keeps the light-front split alive. 
In the next Section, we look at a simplest non-trivial 
quantum mechanics related to light-front $SU(2)$ gluodynamics.\\

{\bf 5. The $P_- = 2\pi/L$ Sector }\\

All this is very nice, but does it really work? 
What are the quantum mechanics appearing in the sectors 
of fixed $P_-$? The real interesting physics is in the 
sectors of large $n\sim L$. But the sectors of low $n$ 
should also contain valuable information. For example, 
they should contain ultraviolet divergences 
from which we can deduce the non-perturbative 
running coupling. This idea that the two-dimensional
quantum mechanics contains dimensional transmutation
was introduced in Ref. \cite{Thorn}.

So, the first test of the approach of Ref. \cite{KMPV} is to try it 
on the sectors of small 
$P_-$. The sector of 
$P_{-} = 0$  was considered in Ref. \cite{KMPV}. 
It is trivial in the sense that there are no 
ultraviolet divergences
in that sector. However, it is quite interesting 
in its own right. The objects one finds in this sector are hybrids of
``fluxes'' known from finite volume equal-time formulation \cite{tHooft},
and non-Abelian plane waves of Coleman \cite{Col}. Considering this
sector allows the 
conclusion that qualitative features of the
infinite volume gluodynamics (like the presence of the mass gap) 
depend on the way the infinite volume limit is taken.

The next in complexity is the sector of $P_- = 2\pi/L$. 
In what follows, we describe the Hamiltonian 
reduced to this sector.\\

{\it 5.0. The Structuring }\\

The description we give is structured: there are three levels.
The first level involves the synthetic quantities constructed from
the fields, like components of the field strength, $F_{kl}$.   
The second level resolves the first expressing the Hamiltonian
in terms of the creation--annihilation operators. The last, third, level
specifies the description to a sector of a fixed number of quanta of the
longitudinal momentum $P_-$. The notations of 
Ref. \cite{KMPV} are used in the 
description.\\

\vspace{.3cm}

{\it 5.1. The Synthetic Level }\\

\vspace{.3cm}
The Hamiltonian consists of the following terms:

\begin{equation}
\label{terms}
H = K + D + F + B.
\end{equation}

$K$ is the kinetic energy of the global angle variable $q$, 
$0 \leq q \leq 1$  \-
($K = -g^2L/(2V(2\pi)^2)(\partial/\partial q)^2$, 
$V$ is the volume of the transverse space).
$D$, $F$, and $B$ are functionally dependent on the transverse
components of the gluon field, $A_k$, and on $q$. $D$ is the 
characteristic light-front term containing the inverse of $D_-$, 
$F$ is the most conventional term---the square of the transverse 
components of the field strength, and $B$ comes from the zero modes. 

Here are the explicit expressions for the terms:

\begin{equation}
\label{D}
 D = \frac{1}{2} \sum_{p \neq 0} 
\bigg| \bigg( \frac{1}{D_-}D_kF_{-k} \bigg)^p\bigg|^2;
\end{equation}

\begin{equation}
\label{F}
F = \frac{1}{4} <F_{lm}|F_{lm}>;
\end{equation}

\begin{equation}
\label{B}
B = -\frac{1}{2} J_k M^{-1} J_k.
\end{equation}
Here $M$ is an operator acting in the space of the zero modes, i.e.,
functions depending only on the transverse coordinates and $x^+$:

\begin{equation}
\label{M}
M = <\chi_0|-D_k^2\chi_0>,
\end{equation}
which is a sum of two terms; 
first of them is just the transverse 
Laplacian, and the second is quadratic in the transverse components $A_k$.
Also, throughout this section, $A_\perp$ is given by
Eq. (\ref{canonical}) with $B_\perp$ set to zero.
The current $J_k$ in Eq. (\ref{B}) is as follows:

\begin{equation}
\label{J}
J_k = <\chi_0|ig \bigg[ A_k,\frac{1}{D_-} D_lF_{-l} \bigg] 
+ D_lF_{lk}>.
\end{equation} 

On the right-hand sides of Eqs. (\ref{D}) - (\ref{B}), 
there is an integration over the transverse coordinates, 
$\int\,\prod_k dx^k$, 
which is not explicitly shown. $J_k$ is a function of the
location in the transverse space.

The inverse of $M$ in Eq. (\ref{B}) is understood in the sense of 
expansion in powers of the fields. It involves the inversion of the 
transverse Laplacian. Here and in what follows, this inversion is 
understood as an operator which annihilates the zero mode of
the function it acts on. \\

{\it 5.2. The Creation--Annihilation Operators Level }\\

Now we need to resolve the synthetic quantities of the previous section in terms of 
the creation--annihilation operators. First we list the relations we will need:
\begin{equation}
\label{creation}
(a^p_k)^\dagger = \sqrt{2p}A^p_k\,, p>0;
\end{equation}
\begin{equation}
\label{conjugation}
A^{-p}_k = (A^p_k)^\dagger;
\end{equation}
\begin{equation}
\label{commutator}
[\chi_{p_1},\chi_{p_2}]=\epsilon(\sigma_1,\sigma_2)
\frac{\chi_{p_1+p_2}}{\sqrt{L}};
\end{equation}
\begin{equation}
\label{aminus}
A_- = \frac{1}{\Delta_\perp}\frac{g}{L}\sum_{p>0}
\sigma_p(a^p_k)^\dagger a^p_k;
\end{equation}
\begin{equation}
\label{p}
p=\frac{2\pi n}{L} + \sigma 
{\rm Dev} \bigg[\frac{2\pi q}{L} + gA_- \bigg].
\end{equation}

In the above relations, the sign $\epsilon$ depending on the two signs of 
deviations of the $p$-eigenvalue off an integer multiple 
of $2\pi/L$ is (i) antisymmetric in its arguments, and (ii) its 
non-trivial values are as follows: 
$\epsilon(+,-) = -$, $\epsilon(0,+) = -$, 
and $\epsilon(0,-) = +$ ($0$ in the arguments of $\epsilon$ appears when 
the corresponding eigenvalue is an
integer multiple of $2\pi/L$); $\Delta_\perp$ is the 
transverse Laplacian,  whose
inversion is understood as above; 
the function ${\rm Dev}[*]$ in Eq. (\ref{p}) 
is by definition as follows:
\begin{equation}
\label{dev}
{\rm Dev}[x] = \min_n \bigg|x -\frac{2\pi n}{L} \bigg|,
\end{equation}   
i.e., it is the magnitude of the deviation of $x$ from its 
nearest integer multiple of 
$2\pi/L$ 
(see also Eq. (\ref{p1})). To avoid misunderstanding, $A_-$ above is 
not a matrix as before. It is a number field, and the right-hand side can be
treated as its definition; it is related to $A_-$ as it was before: 
using the notations of 
Ref. \cite{KMPV}, now $A_- = (\tilde {A}^0_- - \int\,dx^\perp
\tilde {A}^0_- /V)/\sqrt{L}$.

With the equipment of the above relations, we resolve 
the "square root" of the $D$-term as follows:
\begin{eqnarray}
\label{dresolve}
\bigg( \frac{1}{D_-}D_kF_{-k} \bigg)^p 
&  =  & \frac{1}{ip} \bigg(\partial_k (ipA^p_k) - 
    \frac{ig}{\sqrt{L}}
         \sum_{p'\neq 0}ip'\epsilon(\sigma_{p-p'},\sigma_{p'})
A^{p-p'}_kA^{p'}_k \nonumber \\
 & + &  ig\epsilon(\sigma_p,0)A^p_k\partial_kA_- \bigg).
\end{eqnarray}

The rest of the terms in the Hamiltonian are
resolved in the same way.\\

\vspace{.3cm}

{\it 5.3. Specification for a Fixed Value of $P_-$  }\\
\vspace{.3cm}

Specification for a particular sector of fixed longitudinal 
momentum is the most involved part of the description 
of the Hamiltonian. We are not ready to give it for 
the general case of arbitrary fixed $P_-$. We consider the simplest 
non-trivial case.\\

{\it  5.3.1. Specification for $P_- = 2\pi/L$ }\\

First of all, three kinds of excitations are involved: \\
 
(i) the excitation with the 
lowest possible eigenvalue of 
$D_-/i$;  we will call it $a$-excitation, its 
creation--annihilation operators will be denoted by 
$a^\dagger_k$, $a_k$; \\

(ii) the excitation with the next-to-lowest possible eigenvalue 
of $D_-/i$, the $b$-excitation ($b^\dagger_k$, $b_k$); and \\

(iii) the excitation whose $D_-/i$ coincides with the value 
of $P_- = 2\pi/L$, the $c$-excitation ($c^\dagger_k$, $c_k$). \\

So, for example, a $c$-excitation can decay into a pair of $a$ 
and $b$ without breaking the conservation of the longitudinal 
momentum ($a$ carries no longitudinal momentum, and 
$b$ carries the same quantum of the longitudinal 
momentum $2\pi/L$ as $c$ does).  

The reduction of the above Hamiltonian to the sector under 
consideration can be performed by retaining only terms up 
to fourth order in the creation--annihilation 
operators, and containing only the $a-$, $b-$ and $c-$ operators.

The $B$-term above is the most amenable with respect to this reduction. 
So we start the reduction with\\

{\it 5.3.2. $B$-Term Reduction }\\

The $B$-term is a term of the kind  $J_kM^{-1}J_k$, and both 
$J_k$ and $M^{-1}$ are complicated 
(non-polynomial) functionals 
of the creation--annihilation operators. The first step in reduction 
is to notice that the expansion of $J_k$ in the 
creation--annihilation operators starts from the quadratic term. 
Therefore, the leading term in the expansion of the $B$-term 
in powers of the creation--annihilation operators is 
the only one we need for our reduction, and we can 
replace in the leading term $M^{-1}$ by $1/(-\Delta_\perp)$ 
(we recall that there is no ambiguity in the action of 
this operator on a constant function).

The next step is to reveal the leading quadratic contribution to the 
current $J_k$ (see Eq. (\ref{J})).
At the moment, $J_k$ is expressed in terms of $A_k$. $A_k$ 
are expandable in powers of the creation--annihilation operators, 
and the expansion starts from the 
linear term; $J_k$ in turn is expandable in powers of $A_k$, 
and the leading term is quadratic. So, we start from retaining 
the leading term of the expansion of $J_k$ 
in powers of $A_k$:
\begin{equation}
\label{leadingj}
J_k =ig<\chi_0|2\partial_l[A_k,A_l] + [A_l,\partial_kA_l]>.
\end{equation}

All we need to obtain the desired reduction of the $B$-term now is to 
substitute in Eq. (\ref{leadingj}) the expansion 
$A_k=a^\dagger_k\chi_{(0,+)}/\sqrt{2p_a} +
b^\dagger_k\chi_{(1,-)}/\sqrt{2p_b}
+c^\dagger_k\chi_{(1,0)}/\sqrt{2p_c} + h.c.$, 
calculate the commutator involved in Eq. (\ref{leadingj}) 
using Eq. (\ref{commutator}), and compute
the scalar product of the commutator with $\chi_0$ keeping in 
mind that the eigenvectors $\chi$ form the orthonormal set.
Here is the outcome of these manipulations:
\begin{eqnarray}
\label{reducedcurrent}
J_k &=&\frac{2ig}{\sqrt{L}}\partial_l \bigg(\frac{b^\dagger_kb_l-
b^\dagger_lb_k}{p_b} -
\frac{a^\dagger_ka_l-a^\dagger_la_k}{p_a} \bigg) \nonumber\\
& + & \frac{ig}{\sqrt{L}} \bigg(\frac{b^\dagger_l\partial_kb_l - 
(\partial_kb^\dagger_l b_l}{p_b} -
\frac{a^\dagger_l\partial_ka_l - (\partial_ka^\dagger_l)a_l}{p_a} \bigg).
\end{eqnarray}
Here 
$p_a = {\rm Dev}[2\pi q/L]$, $p_b = 2\pi/L - {\rm Dev}[2\pi q/L]$, 
and $p_c = p_a + p_b$.

If we substitute the above expression of the $J_k$ in the $B$-term, 
it will generate a number of contributions which can be classified 
as $b$-$b$ interaction, $a$-$a$ interaction, and $a$-$b$ interaction. 
In our sector, we need only the $a$-$b$ 
contribution (because we have only one $a$-excitation, and only 
one $b$-excitation). Retaining the $a$-$b$ contribution we obtain
\begin{equation}
\label{breduced}
B = \frac{g^2}{L}\frac{1}{p_a p_b} j_k(b)\frac{1}{\Delta_\perp}j_k(a),
\end{equation} 
where 
\begin{equation}
\label{jreduced}
j_k(a) = 2\partial_l(a^\dagger_ka_l-a^\dagger_la_k) + 
(a^\dagger_l\partial_ka_l - (\partial_ka^\dagger_l)a_l),
\end{equation}
$j_k(b)$ is obtained  from $j_k(a)$ by the substitution $a\rightarrow b$.
The last thing to notice is that the ugly non-smooth 
${\rm Dev}$-function featuring the expressions for $p_a$, $p_b$ 
can be dropped out of Eq. (\ref{breduced}). To see how it works, 
consider a small value of $q$ and observe that here 
$p_ap_b = (2\pi/L)^2q(1-q)$ which is symmetric with respect 
to the reflection $q\rightarrow 1-q$. Therefore, 
the exact expression with ${\rm Dev}$, which works also for 
the values of $q$ exceeding $1/2$, is identical in this case 
with the above naive representation 
without  ${\rm Dev}$, which in general holds only at $q<1/2$.

Therefore, our final expression for the reduced $B$-term is as follows:
\begin{equation}
\label{finalbreduced}
B=\frac{g^2}{(2\pi)^2}\frac{L}{q(1-q)}j_k(b)\frac{1}{\Delta_\perp}j_k(a).
\end{equation}
\\


{\it 5.3.3. Reduction of the $D$- and $F$-terms }\\

To reduce the $D$- and $F$- terms we notice that they both are 
infinite sums over the spectrum of $D_-/i$. Only finite number 
of terms of these sums contribute 
when the action of the Hamiltonian is reduced to the sector 
$P_- = 2\pi/L$. In fact, 
an inspection reveals that only the terms whose eigenvalues $p$ satisfy 
$p\leq P_-$ contribute. Therefore, the $D$-term contribution is in fact
$|D^a|^2+|D^b|^2+|D^c|^2$, where, for example, $D^a = ((1/D_-)D_kF_{-k})^a$; 
and the $F$-term contribution is 
$(|F_{lm}^0|^2+|F_{lm}^a|^2+|F_{lm}^b|^2+|F_{lm}^c|^2)/2$, where, for 
example, $F_{lm}^0 = -ig([A_l,A_m])^0$.

When the transverse components $A_k$ are replaced in those expressions 
by the sum of the creation--annihilation operators divided by 
the square roots of the corresponding eigenvalues, 
non-polynomial functions of the creation--annihilation 
operators appear, because the square roots of the eigenvalues 
downstairs contain $A_-$, which is quadratic in 
the creation--annihilation operators (see Eq. 
(\ref{aminus})). For our reduction, we need to expand these 
functions in $A_-$ and to retain only the linear terms in $A_-$. 
A Characteristic example is as follows:
\begin{equation}
\label{example}
\frac{1}{\sqrt{2p}} \simeq \frac{1}{\sqrt{2\bar{p}}}
\bigg( 1 -
\frac{g\sigma\epsilon(q)}{2\bar{p}}A_- \bigg),
\end{equation}
where $\bar{p}$ is $p$ at zero $A_-$, $\sigma$ is the sign 
of the deviation of $p$ off the multiple integer of $2\pi/L$, 
and $\epsilon(q) = +1$ when $q<1/2$ and $-1$ otherwise. 

When Eq. (\ref{aminus}) is used to express $A_-$ in terms of the 
creation--annihilation operators, a characteristic 
"potential" generated by the $a$- or $b$-
charges enters the formulas. We introduce a dedicated notation for it:
\begin{equation}
\label{potential}
V_{a}=\frac{1}{-\Delta_\perp}a^\dagger_ka_k,\, 
V_{b}=\frac{1}{-\Delta_\perp}b^\dagger_kb_k.
\end{equation}
Also, we will use another useful notation:
\begin{equation}
\label{cal}
{\cal A}_k = \frac{a_k}{\sqrt{2p_a}},
\end{equation}
and similarly for other creation--annihilation operators (for example, 
${\cal C}^\dagger_k = c^\dagger_k/\sqrt{2p_c}$). 

With these notations, the reduced expressions for 
$D^a$, $D^b$, \ldots, $F^c$ 
are as follows (by reduced we mean that 
some terms have  been omitted 
because they are vanishing in the sector under consideration):
\begin{equation}
\label{da}
D^a = \bigg(1-\frac{g^2\epsilon(q)}{2Lp_a}V_b \bigg)\partial_k
{\cal A}_k^\dagger + \bigg(1+\frac{\epsilon(q)}{2}\bigg)
\frac{g^2}{Lp_a}
(\partial_kV_b){\cal A}_k^\dagger
+\frac{-ig}{\sqrt{L}}\frac{p_c+p_b}{p_a}{\cal C}_k^\dagger{\cal B}_k;
\end{equation}
\begin{equation}
\label{dc}
D^c = \partial_k{\cal C}_k^\dagger + 
\frac{ig}{\sqrt{L}}\frac{p_b-p_a}{p_c}{\cal B}_k^\dagger{\cal A}_k^\dagger;
\end{equation}
\begin{equation}
\label{f0}
F^0_{lm} = \frac{-ig}{\sqrt{L}}({\cal A}_l{\cal A}^\dagger_m - {\cal A}_m
{\cal A}^\dagger_l)
+ \frac{ig}{\sqrt{L}}({\cal B}_l{\cal B}^\dagger_m - {\cal B}_m
{\cal B}^\dagger_l);
\end{equation}
\begin{eqnarray}
\label{fa}
F^a_{lm} & = & \partial_l \bigg({\cal A}_m^\dagger 
\bigg(1-\frac{g^2V_b\epsilon(q)}{2Lp_a}\bigg)\bigg)-
\partial_m \bigg({\cal A}_l^\dagger 
\bigg(1-\frac{g^2V_b\epsilon(q)}{2Lp_a}\bigg)\bigg) \nonumber \\
& + & \frac{-ig}{\sqrt{L}}({\cal B}_l{\cal C}_m^\dagger 
- {\cal B}_m {\cal C}_l^\dagger);
\end{eqnarray}
\begin{equation}
\label{fc}
F^c_{lm} = \partial_l{\cal C}_m^\dagger - \partial_m{\cal C}_l^\dagger -
\frac{ig}{\sqrt{L}}({\cal B}^\dagger_l{\cal A}^\dagger_m - 
{\cal B}^\dagger_m{\cal A}^\dagger_l).
\end{equation}
The two missing expressions for $D^b$ and $F^b_{lm}$ are obtained from the 
expressions for $D^a$ and $F^a_{lm}$ by the substitutions ${\cal A}\rightarrow 
{\cal B}$, $a\rightarrow b$, and
$-i\rightarrow i$ for the imaginary unit.

With these expressions at our disposal, the reduced Hamiltonian is derived by 
taking the sum of 
their magnitudes squared, and by omitting the excessive terms.\\

{\it 5.3.4. The three-gluon vertex}\\

The terms bilinear in the creation-annihilation operators are easily 
obtainable;  
they are as they should be, 
$a^\dagger_k(-\Delta_\perp)/(2p_a)a_k + 
b^\dagger_k(-\Delta_\perp)/(2p_b)b_k +
c^\dagger_k(-\Delta_\perp)/(2p_c)c_k$.

The next in complexity is the 
three-gluon vertex describing the decay 
of a $c$-excitation into 
a pair of $a$- and $b$- excitations. We will express it in terms 
of the Fourier modes of the above
operators, ${\cal A}^\dagger_k(x) \equiv 
\sum_{k^{a}}(\tilde{A}^\dagger_k(k^a)\exp{ik^ax})/\sqrt{V}$, etc., 
where $x$ is the location in the transverse space, 
$k^a$ is the transverse momentum of the 
$a$-excitation, and the $V$ is the volume of the transverse space. 
The term of the Hamiltonian we are looking for is 
representable as follows:
\begin{equation}
\label{fourier}
G_3 = \sum_{k^a,k^b,k^c}\delta(k^c-k^a-k^b)\tilde{A}^\dagger_a(k^a)
\tilde{B}^\dagger_b(k^b)
\tilde{C}_c(k^c)V_{abc}(k^a,k^b,k^c) + h.c.
\end{equation}
Now our task is to get an expression for $V_{abc}$ 
(summation over the transverse vectorial indices $abc$ is implied above).

A calculation along the above lines gives
\begin{equation}
\label{g3}
V_{abc}(k^a, k^b, k^c) = 
\frac{2g}{\sqrt{LV}} \bigg[\frac{\delta_{bc}}{p_a}(k^b_ap_a - k^a_ap_b)
+\frac{\delta_{ac}}{p_b}(k^b_bp_a - k^a_bp_b) - 
\frac{\delta_{ab}}{p_c}(k^b_cp_a - k^a_cp_b) \bigg].
\end{equation}
We have a loose notation in the above formula: e.g., the subscript $a$ 
denotes both vectorial index of the $a$-excitation, 
and the label on the $D_-/i$-eigenvalue. To avoid misunderstanding, 
it is a vectorial index when it hangs on the transverse momenta, 
or on a Kronecker's delta-symbol.

The only term we are to determine in the reduced 
Hamiltonian $H_{red} = K + G_2  + G_3 +G_4$ is the $G_4$-term.
Partly we know it, because we specified above reduction of the $B$-term,
which gives a contribution to $G_4$.There are quite a number 
of terms in $G_4$, and we will not write them down explicitly. 
We hope that the general pattern is 
clear, and the reader can recover the rest of the terms in $G_4$.

Comparison with Ref. \cite{Thorn} shows that the dimensional transmutation 
is implied by this Hamiltonian. The crucial check of 
the whole construction can be given by a calculation of the numerical 
coefficient by the leading inverse logarithm of the ultraviolet 
cut-off in the running coupling, because its value is known from 
conventional perturbation theory. This calculation is in progress. \\

{\bf 6. Conclusion and Outlook}\\

We described a promising approach to a non-perturbative
description of non-Abelian gauge theories. 
It results from previous efforts to answer the challenge 
of the light-front quantisation of the gauge theories 
(see Ref. \cite{Brodsky}), and from recent 
analysis of the light-front formulation \cite{KMPV}. 
The first result we expect 
from this approach is an alternative non-perturbative 
definition of the running coupling that can be obtained from the 
quantum mechanics in the sector $P_-=2\pi/L$.

In more general terms, we are at 
the very beginning of a long road: 
we need to generalise  to $SU(N)$, to include fermions, and, 
the most interesting, to go to the sectors of large $P_-$. 

We note also that the finite volume light-front formulation 
may play an important role for string theory,
where one has to quantise  a compactified theory.
Since we have obtained a light-front formulation 
without a gauge fixing 
and in finite volume our results can stimulate 
a deeper understanding 
of a relation with novel
M-theory developments \cite{Susskind}.

It is too early to make a conclusion about the 
approach we presented, but we believe in its 
promising future.

{\bf Acknowledgements.}
GBP thanks the Organising Committee of the XXXIV PNPI Winter School
for kind hospitality and support. 
This work was supported  in part
by the Russian Foundation 
for Basic Research,  the NATO Science Programme,
the U.S. Department of Energy and the U.S. National Science Foundation.

\end{document}